# MESHAT: MONITORING AND EXPERIENCE SHARING TOOL FOR PROJECT-BASED LEARNING

Christine Michel *, Élise Garrot-Lavoué *, **
* University of Lyon, INSA Lyon, LIESP laboratory
Bâtiment Léonard de Vinci, 21 avenue Jean Capelle, F-69621 Villeurbanne, France
** University of Lyon, University Jean Moulin Lyon 3, IAE, Centre de recherche MAGELLAN
6 cours Albert Thomas BP 8242 - 69355 Lyon Cedex 08
Christine.Michel@insa-lyon.fr, Elise.Lavoue@univ-lyon3.fr

**ABSTRACT**

Our work aims at studying tools offered to learners and tutors involved in face-to-face or blended project-based learning activities. To understand better the needs and expectations of each actor, we are especially interested in the specific case of project management training. The results of a course observation show that the lack of monitoring and expertise transfer tools involves important dysfunctions in the course organisation and therefore dissatisfaction for tutors and students (in particular about the acquisition of knowledge and expertise). So as to solve this problem, we propose a personalised platform (according to the actor: project group, student or tutor) which gives information to monitor activities and supports the acquisition and transfer of expertise. This platform is meant for the complex educational context of project-based learning. Indeed, as for the majority of project-based learning activities, the articulation conceptualisation-experiment is an important part of the process. The originality of our approach relies on also supporting the articulation between action (experiment or conceptualisation) and reflection. This approach so improves the acquisition of complex skills (e.g. management, communication and collaboration), which requires a behavioural evolution. We aim at making the students become able 'to learn to learn' and evolve according to contexts. We facilitate their ability to have a critical analysis of their actions according to the situations they encounter.

## 1. INTRODUCTION

Project-based learning is often applied in the case of complex learning (i.e. which aims at making learners acquire various linked skills or develop their behaviours). In comparison to traditional learning, this type of learning relies on co-development, collective responsibility and co-operation (Huber, 2005). Learners are the principal actors of their learning. A significant enrichment arises from their activity, both for them and all the other learners. A consequence of this approach is the segmentation of the class into sub-grouped projects, monitored by tutors. We generally observe that the coordination and harmonisation of tutors' activities are extremely difficult to operate when each group works autonomously, on different subjects and in real and varied environments (for example enterprises). It is even more difficult when the project is conducted over a long period (more than four weeks). In this context, the perception of individual's and group's activity is also very difficult, especially if no technical support for information and communication is used. Finally, the implementation of project-based learning in engineering schools, universities or professional training do not benefit from all its capacities (Thomas & Mengel, 2008). Indeed, this learning should implement an educational model based on the Kolb's cycle (Cortez *et al.*, 2008), composed of four phases (personal experience, reflexive observation, conceptualisation, experiment). But it is often *action* (via the articulation conceptualisation-experiment) which is favoured to the detriment of *reflection* and of *personal experience* (Thomas & Mengel, 2008).

To understand better the type of tool necessary to improve these trainings, we have studied a project management training course (Michel & Prévot, 2009). This course is supported by a rich and complex organisation, especially for tutors that we detail in part 2. We have used KM methods to identify all the problems encountered by students and tutors. We mainly identify three problems: (1) difficulties for students to acquire some skills (e.g. project management organisation, use of monitoring tools and group work) and autonomy, (2) lack of information so that tutors can monitor and evaluate students individually and by group, and (3) lack of tutors' communication and coordination so that they develop their expertise, knowledge and competences. In part 3, we study existing tools which can help to solve these problems, especially monitoring and experience sharing tools. We then observe that no existing tool could solve all these problems on its own. We thus propose a new tool named MEShaT (for **M**onitoring and **E**xperience **Sha**ring **T**ool). We finally conclude by all the futures directions offered by this work.

## 2. CASE STUDY: A PROJECT MANAGEMENT TRAINING COURSE

### 2.1 The course organisation

The course is composed of a theoretical presentation of the principles and methods of project management and their practical application to a project (called 'PCo' for 'Collective Project') carried out by groups (12 groups of 8 students which answer to different industrial needs). Envisaged by Patrick Prévôt (Michel & Prévot, 2009), the project management course lasts six months and corresponds to an investment of approximately 3000 students' working hours per project. The instructional objectives are to acquire hard competences (e.g. knowing how to plan the project (Gantt's chart), to organise the project management, to manage resources, to control quality) and soft competences (e.g. social competences of collaboration and communication, empathy, consideration of the others, leadership). The pedagogical team (see figure 1) is composed of 24 tutors (a technical and a management tutor per group), 2 managers (technical and management) in charge of the coordination of the technical and management tutors' activities, 1 teacher who presents the theoretical concepts and 1 director responsible for the organisation of the training of all groups.

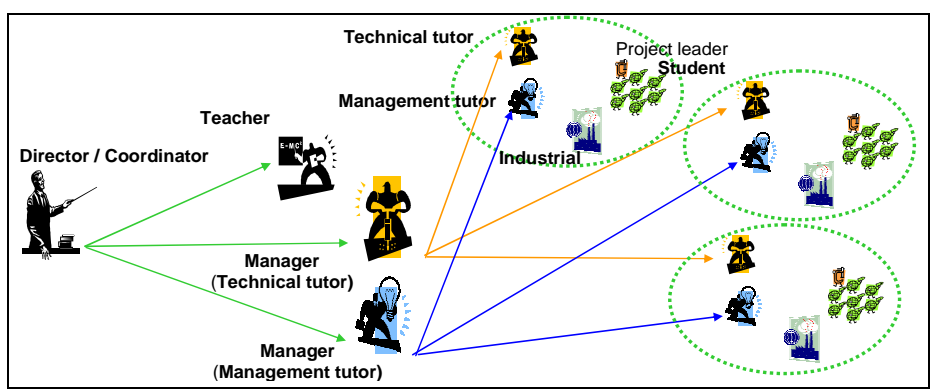

Figure 1. Pedagogical team and course organisation

The project is composed of four phases:
- (1) November: answer to the call for tender (formalisation of the client's requirements).
- (2) December: elaboration of a master plan (means, tools and organisation of the team project), definition of tools to drive the project (dashboard) and rules to test the deliverables quality (rules of receipt).
- (3) January to March: development of a product or a study.
- (4) Until mid-April: delivery of a technical report which describes the product and management report (a project closure report which is an analysis, from the student's point of view, of the flow and problems of the project). The project is closed by one dramatised presentation in front of all the actors of the project.

The tutors play various roles which depend on the type of skills that students have to acquire (Michel, 2009). According to Garrot's taxonomy (Garrot *et al.*, 2009), for the acquisition of soft skills, tutors are social catalysts (create a friendly environment to incite students to participate), intellectual catalysts (ask questions and incite students to discuss, to criticise), 'individualisers' (help every student to overcome the difficulties, to estimate his/her needs, difficulties, preferences) and 'autonomisers' (help students to regulate their learning and to acquire autonomy). For the acquisition of hard skills, tutors are relational coaches (help students to learn how to work in group and to become a leader), educationalists (redirect groups' activities in a productive way, clarify points of methodology, supply resources), content experts (answer to questions on the course contents), evaluators (evaluate students and groups' productions and participation) and 'qualimetrors' (measure and give feedback on the quality of the course).

Tutors monitor a unique and non reproducible project. They work with students most of time in face-to-face and no organisation, communication or capitalisation tool is proposed. For example, no specific tool is currently proposed to the tutors for the monitoring of students' activities or for their evaluation. The appreciation of students' activity is made in an implicit way, according to the number and the quality of face-to-face students–tutor interactions. In terms of communication and coordination, each tutor works individually with his/her group and does not communicate much with the other tutor of the same group (management or technical) in order to have a complete vision of the group's activity.

## 2.2 The observed problems

The observation methodology is adapted from the MASK (Method for Knowledge System Management) method (Benmahamed1 *et al*, 2005), completely described in (Michel & Prévot, 2009). The observation data are various experience feedbacks from students and tutors and were collected by 62 students in the 5th year of the engineer school. The observed students are 23 males and 18 females and are between 22 to 25 years old. 38 of them have carried out the project management course the previous year, 3 of them are currently 'project leader' in a project. Observation consists in direct feedbacks (called REX) made by interview of the course director, of 6 tutors and of 3 students currently 'project leader' and by self-observation for the other 38 students. Indirect feedbacks are based on various groups' experience and analyses expressed in there 'management report', which is one of the projects deliverables. 24 management reports have been considered (each one relating to the experience of a group). Each identified problem has been described on a RISE (Reuse, Improve and Share Experiment) card. We have observed 36 different types of problems.

The majority of cards (57%) relate to a problem with the *management of the team work by the team itself*. More precisely, 29% relate to a lack of project management skills, 18% relate to difficulties working in group, 10% relate to problems with some students who think they are not responsible enough. 31% of the problems concern *tutors' activity* and impact on *teaching organisation of the project*. Indeed, 13% concern a lack of coherence, coordination and communication between tutors, which involve problems of information diffusion. For example, the instructions given to the project groups were described as ambiguous or contradictory. About 5% concern a lack of communication between tutors and students or a lack of presence of some tutors. 13% concern a lack of information for tutors on the teaching objectives or on the knowledge and skills they have to teach to students. Indeed, students feel them alone when they have to learn using some tools or when they have to apply theoretical project management concepts. Students sometime do not understand the role tutors play and the help they can bring them. Moreover, 8% of the problems concern failure in the *teaching design of the course* (not enough time to work, a not adapted calendar and a too short timing for the deliverables). Finally, many groups and tutors express a same problem concerning the *monitoring of individuals' or groups' activity and students evaluation* (4% of the problems). The students express a feeling of injustice concerning the individual evaluation because the notation is the same for all the members of a project (with about + or -2 points according to their investment), even if the students involved more or less than the others. All the tutors also express their difficulties to evaluate the students individually. These difficulties are explained by the intuitive and tacit character of the evaluations, by the lack of traceability of students' actions, and by the lack of discussion with their colleagues.

It is possible to partially solve problems concerning the course design and the course organisation by changing the timing and the teacher's and coordinator's responsibilities. Nevertheless many problems remain and most of them are directly or indirectly bound to tutors' activity. That is why we aim at helping tutors, on the one hand, to monitor and to evaluate the students and the groups and, on the other hand, to exchange information, coordinate and develop their skills and expertise. Although the pedagogical context is not distance learning, we hope to benefit from using tools to support this activity. In next part, we study CSCL and Web 2.0 tools which are suitable for our case. We focus on monitoring tools and expertise sharing tools.

## 3. TOOLS TO SUPPORT LEARNING ACTIVITIES

In this part, we detail existing tools to help tutors to monitor students' activities and to communicate with the other tutors. We study how these tools can help tutors and solve the problems identified in the previous part. We finally show that none of them answer to all the needs for which we develop our own tool.

### 3.1 Monitoring tools

Many tools have been developed to support tutors in the monitoring of distant and synchronous students' individual activities. ESSAIM (Després, 2003) gives a global view of a student's progress in the course and tutors have a perception of the activity with referring to the path, the actions and the productions of each student. FORMID (Guéraud & Cagnat, 2006) offers a tutor interface with a global view of a class during a session (e.g. students' login, their progress in the course) or a zoom in on a precise course stage (successfully validated or not by the class, by a student or by a group of students so as to identify their difficulties). These tools work in a synchronous environment with automatically generated tracks. They are thus only meant for tutors and do not offer the possibility to students to regulate their learning for a long period. Furthermore, they are not meant for asynchronous learning situations for which tutors need information on learners' activities on a long period.

Other tools are meant for helping tutors to monitor asynchronous activities and allow to go with the learners towards their autonomy or to regulate their learning by determine themselves the state of their progress in the course. Croisières (Gueye, 2005) offers services which support individually learners in their learning progress and assist them in autonomy situation. Learners select their learning activities according to their objectives and learning strategies. Reflet (Després & Coffinet, 2004) is a tool meant for showing the state of progress of a student or a class. It supplies information to the tutors who monitor the students in distance training and to the students who have a feedback on their progress with regard to the learning objectives and the other students. Learners determine their state of progress in the course with regard to the tasks they have to carry out and tutors can deny learners the validation of some of their tasks.

There also are tools to monitor the activities of groups, not simply individuals. SIGFAD (Mbala *et al.*, 2005) offers a support for actors' interactions in restricted groups (8-15 persons) in distance learning. It helps tutors to hold the groups, to boost them and indeed to conduct well the course. The interaction statistics allow to model and to show the collaboration into groups, to estimate the group's life and evolution. SIGFAD supplies three main categories of estimations: at the level of the group (present, absent or still persons, the state of the group with regard to the realisation of the activities), at the level of individuals (their productivity in term of realisation of the activities and their sociability which indicates their level of communication with the other members of the group) and at the level of the activity (level of realisation of an activity by all the participants). TACSI (Laperrousaz *et al.*, 2005) offers more specifically a perception of the individual learners' activity into the activity of their group. It distinguishes the perception of learners' activity in an individual task (individual productions), the perception of learners' activity in a collective task (their contributions in the collective activities and their contributions to the discussions) and the perception of learners' situation in the group dynamics (social behaviour and sociometric status). The LCC (Learning to Collaborate by Collaborating) collaborative activity software (Cortez *et. al*, 2008) is used for teaching and measuring teamwork skills using technologically supported face-to-face collaborative activities. LCC allows measuring seven variables : first ones measure the *Activity score* (i.e. the group's efficiency in performing the task assigned), last ones measure *Teamwork variables (*corresponding to core components (skills) of teamwork like team orientation (TO), team leadership (TL), monitoring (MO), feedback (FE), back-up (BA) and coordination (CO)). Communication has not been included in the measurable variables.

The individual and collective indicators for the monitoring of learners and project groups offered by these tools are relatively well adapted to our context. We especially adopt those proposed within the LCC framework (Cortez *et al.*, 2008) for the development of our own monitoring tool. However, the course which interests us does not use instrumented activity and does not thus allow using automatically collected tracks of learners' activity. That is why we have to think about other ways of collecting information on their activities.

The tools which help learners to acquire autonomy incite them to evaluate their progress in the course, according to the tasks they have achieved and those they have to achieve. However, these tools are not adapted because they do not help learners to build an individual reflection neither on the relevance of the knowledge they acquire and the modalities of this acquisition nor on their behavioural changes. These self-regulatory processes are individual and mainly result from the activities carried out with the tutors. We think useful (Michel, 2009) to support these processes by using a metacognitive tool (Azevedo, 2007) which takes into account learners' point of view of: *cognition* (e.g. activating prior knowledge, planning, creating sub-goals, learning strategies), *metacognition* (e.g. feeling of knowing, judgment of learning, content evaluation), *motivation* (e.g. self-efficacy, task value, interest, effort) and *behaviour* (e.g. engaging in help-seeking behaviour, modifying learning conditions, handling task difficulties and demands).

All the tools studied in this part are exclusively centred on learners' activity and help neither learners nor tutors to have reflections on their activity. In our context, in which the roles played by tutors are extremely varied, it is essential to have a base structuring this reflection. For example, Berggren & Söderlund (2008) propose to use a 'learning contract' defined as *'a number of fairly simple questions, such as: What do I want to learn? How will I learn this? Who can give support? When can I start? How will I know that I have learned? How will others realize that I have learned?'*. This contract could be useful not only for students but also for tutors.

Furthermore, all the tools do not brought help to tutors to understand or interpret what they observe. They supply useful information for tutors but these information are rather quantitative than qualitative and do not thus allow to evaluate the quality of the contributions or productions, or to explain learners' behaviour neither individually nor inside the group. These tools can be useful for tutors only if they know how to use it, how to interpret the supplied information and how to react effectively and in an adapted way. Finally, these tools address every tutor individually and do not allow them to coordinate at the level of the monitoring of a same project group and to exchange on their activity so as to acquire more expertise. That is why we study in next part the tools which support exchanges between tutors to bring them to help each other and to develop their skills.

## 3.2 Experience sharing tools

The results of a previous study (Michel *et al.*, 2007) about tools supplied to the tutors show that they do not have adapted tools to exchange or formalise their experience, as allowed for example by Knowledge Based Systems (KBS) or experience booklets (Kamsu Foguem *et al.*, 2008). Furthermore, we observed that tutors are rather structured in a hierarchical way within the organisation and do not have coordination tools or dedicated spaces for meeting between peers.

To compensate for a lack of training and formal help, Communities of Practice (CoPs) of tutors emerge. Web technologies (e.g. forums, blogs, wikis) have allowed the emergence of online CoPs (Cuthell, 2008; Pashnyak & Dennen, 2007). CoPs gather tutors together in an informal way because of the fact that they have common practices, interests and purposes (i.e. to share ideas and experiences, build common tools, and develop relations between peers). Members exchange information, help each other to develop their skills and expertise and solve problems in an innovative way. They develop a community identity around shared knowledge, common approaches and established practices and create a shared directory of common resources (Wenger, 1998; Garrot-Lavoué, 2009). The use of technology does allow the accumulation of exchanges, but they are relatively unstructured and not contextualised. Web tools such as blogs, mailing lists, chat and email, allow discussion without building concrete knowledge (only forums bring a slightly higher degree of explicit emergence, thanks to the spatial representation as discussion threads which highlights relations between messages).

Numerous works aim at answering the question by supplying tutors with tools to support specific activities. Some tools work through member participation and sociability, for example by offering a virtual 'home' like the Tapped In environment (Schlager & Fusco, 2004), others by supporting collaboration between members like CoPe_it! (Karacapilidis & Tzagarakis, 2007). Other tools favour the creation of contextualised resources and contextual search facilities such the learning environment doceNet (Brito Mírian *et al,*. 2006). However, all these environments either favour sociability (engaging members to participate) to the detriment of the reification of the produced resources, or they favour the accumulation and indexation of contextualised resources, but to the detriment of sociability and member participation.

We have developed the TE-Cap platform (Garrot-Lavoué, 2009) so as to support a good structuralisation of the information without decreasing the member participation (for example communication). Indeed, the tutors have discussions by way of contextualised forums: they associate tags with the discussions to describe the context. These tags are subjects of a tutoring taxonomy, showed in an interactive and evolutionary way (the tutors can propose new subjects for the taxonomy). This platform, associated with a monitoring tool, could answers our needs of knowledge and skills acquisition and capitalisation about the realisation of tutors' activity and about the use of the monitoring tools.

## 4. A PLATFORM FOR TUTORS AND STUDENTS

We have designed a customised platform called MEShaT (see figure 2). It proposes different interfaces according to the learning actor: a project group, a student or a tutor. Every interface consists of (1) a monitoring tool (on the form of a dashboard) which helps the concerned actor to have a global view of his/her activity and (2) a publication tool which allows spreading his/her experience.

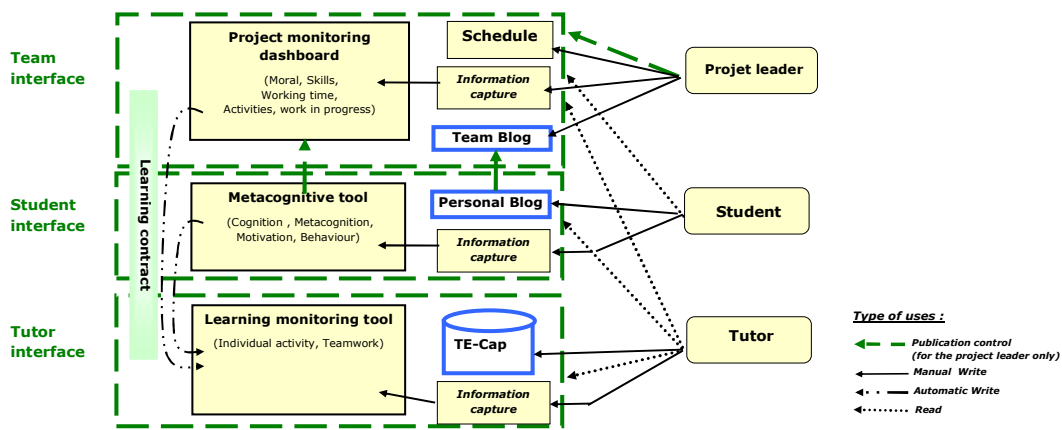

Figure 2. MEShaT : Monitoring and Experience Sharing Tool for project-based learning

Three dashboards are offered; two for students (one to monitor the progress of their project and the other one to monitor their own learning process) and one for tutors (to monitor students' and groups' activities and students' learning). The *project monitoring dashboard* is meant for the project group and shows various indicators: the group's frame of mind, the skills to build for the project realisation, the working time of each member, tasks to realise, the deliverables to produce, the delays, etc. The *metacognitive tool* is meant for a student and shows reflexive indicators. It aims at improving the learning in the case of complex domain or when a behavioural evolution is expected. According to Azevedo (2007) (see part 3.1), we consider that it is important to reflect the students' cognition, metacognition, motivation and behaviour. The *learning monitoring tool* is meant for tutor and shows information on the individual students' activity and the groups' activity thanks to indicators such as the group orientation, leadership, monitoring, feedback and coordination (Cortez *et al.,* 2008).

The publication tools are *blogs* and *TE-Cap*. Blogs (one per student and one per group) are spaces where students can freely describe for example the realisation contexts of their actions and their frame of mind. These blogs help the group members and the tutors to understand the project context, to explain the value of some indicators (as delays or the group's frame of mind) and so to anticipate or to solve more quickly the problems. TE-Cap is offered to tutors to allow the emergence of a Community of Practice composed of all the tutors who monitor a project. The indexation model is built on three main subjects, corresponding to the different types of expertise required for tutors: (1) their roles and tasks, (2) the project calendar (so as to coordinate) and (3) the specific progress of every group. By exchanging, tutors will acquire expertise on their roles and knowledge on their application ground. TE-Cap can be considered as an expertise transfer tool.

A fixed section shows information accessible by all the actors: the *learning contract* (see part 3.1). The learning contract (Berggren & Söderlund, 2008) helps every actor (tutor and student) to evaluate him/herself in regards to the educational model and to better reach his/her objectives.

We detail the modalities of use (represented by arrows on figure 2). The information on the dashboards can be modified by their owner(s) and are not visible for everybody. Students can modify their blog and their individual dashboard by means of a data entry interface. The groups' dashboard is updated by the project leader, using individual information. Leaders confirm the data and decide what is published on the blog. The tutors' dashboard is directly updated by them and automatically updated according to the information entry on the groups' and students' interfaces. Tutors also contribute directly to the CoP. Tutors have access to the groups' and students' interfaces. The project leaders have not access to the individual dashboards of their group members. The learning contract cannot be modified during the course progress. It is updated at the end of the project, according to the events which were related on blogs and on TE-Cap.

MEShaT is meant for the complex educational context of project-based learning, using the Kolb's learning process. The metacognitive tool, the blogs, TE-Cap and the learning contract, favour the *reflection* and *personal experience* phases of the Kolb's cycle, the monitoring tools help *action* phases (conceptualisation and experiment). Moreover, MEShaT solves some of the problems identified in section 2.2. Monitoring tools and blogs facilitate the group work, the group cohesion, the professionalism of students by making more tangible the consequences of their acts and by informing them. Metacognitive tool and blogs help the students to acquire knowledge and reinforce their motivation (by a better understanding of what they have to do and why they do it). If these phenomena do not naturally appear, tools will help the tutors to make them emerge. Indeed, MEShat reinforces the tutor-student link by allowing a continuous monitoring of the knowledge acquisition process. It also helps tutors to assume some of their roles, like their roles of relational coach and social catalyst (concerning the group work or the leadership), their role of intellectual catalyst (by asking precise and conceptualised questions to incite students to discuss or ask critical questions) and their roles of expert and pedagogue. Moreover, the association of Te-Cap with the learning contract offers to tutors a space for refining or developing their expertise.

## 5.  CONCLUSION AND FUTURES DIRECTIONS

Our work aims at studying the monitoring and expertise transfer tools proposed to tutors and students involved in face-to-face or blended project-based learning activities. To understand better the needs and expectations of each actor, we are especially interested in the case of project management training. Indeed, this type of learning is complex since it has for objective the acquisitions of soft and hard knowledge and relies on rich and varied social organisations. In the first part of this article we described a course which is based on these principles. The lack of monitoring and expertise transfer tools involves important dysfunctions in the course organisation and therefore dissatisfaction for tutors and students (in particular about the acquisition of knowledge and expertise). The study of existing tools highlights two points: (1) there is no tool which help both tutors and students, (2) there are not clear strategies proposed to acquire, transfer and capitalise the actors' experience. Indeed, studied tools do not offer metacognitive functions, formal or informal publication tools (as knowledge books or blogs) and tools to support Communities of Practice.

So as to solve this problem, we propose to associate personalised monitoring tools (one for the project group, one for the student and one for the tutor) with tools for the transfer of experience and the acquisition of knowledge. Regarding the monitoring: the 'Team feedback' is a dashboard for the project management, the 'Student feedback' is a metacognitive tool and the 'Tutor feedback' is a monitoring tool for individuals' and groups' activity. The tool for the acquisition of knowledge considers two types of knowledge: the acquired experience is formalised in a kind of knowledge book called 'learning contract', the experience being acquired is revealed and capitalised in blogs (for students and project groups) and within a CoP supported by TE-Cap (for tutors). We describe their articulation in a platform: MEShaT. This platform is dedicated to project management education but can also be used to support different type of face-to-face project-based learning activities. Indeed, all the phases of the Kolb's cycle are well taken into account. Furthermore, it supports the acquisition of various experiences: those of the individuals (students and tutors) and those of the social organisations (project group, CoP of tutors). Our future work will consist in testing this platform on a long time so as to validate experimentally our hypotheses. We will also observe how the actors (students, tutors and course designer) appropriate this type of technologies and how these last ones participate in the redefinition of their roles.